\documentclass[12pt,a4paper,oneside]{article}

\usepackage[margin=1in]{geometry}
\usepackage{newtxtext,newtxmath}

\usepackage{graphicx}	% Including figure files
\usepackage{amsmath}	% Advanced maths commands
\usepackage{amssymb}	% Extra maths symbols
\usepackage{multicol}        % Multi-column entries in tables
\usepackage{natbib}
\bibpunct{(}{)}{;}{a}{}{,} % to follow the A&A style
\usepackage{tablefootnote}
\usepackage{hyperref}
\usepackage{url}

\usepackage{ae,aecompl}
\usepackage{graphicx}	% Including figur
\usepackage[affil-it]{authblk} 
\usepackage{etoolbox}
\usepackage{lmodern}
\usepackage{threeparttable}

\def\fdeg{\hbox{$.\!\!^\circ$}}

\def\farcsec{\hbox{$.\!\!^{\prime\prime}$}}
\def\sbmag{mag~arcsec$^{-2}$}
\def\deg{$^\circ$}
\def\arcmin{\mbox{$^\prime$}}%
\def\arcsec{\hbox{$^{\prime\prime}$}}

\title{CASTLE: performances and science cases}

\author[1]{S. Lombardo}
\author[2]{F. Prada}
\author[1]{E. Hugot}
\author[1]{S. Basa}
\author[6]{J. M. Bautista}
\author[1]{S. Boissier}
\author[1]{A. Boselli}
\author[1]{\\A. Bosma}
\author[3]{J. C. Cuillandre}
\author[4]{P. A. Duc}
\author[1]{M. Ferrari}
\author[1]{N. Grosso}
\author[5]{L. Izzo}
\author[1]{\\K. Joaquina}
\author[1]{Junais}
\author[6]{J. Koda}
\author[7]{A. Lamberts}
\author[1]{G.~R. Lemaitre}
\author[1]{A. Longobardi}
\author[2]{\\D. Mart{\'\i}nez-Delgado}
\author[1]{E. Muslimov}
\author[2]{J. L. Ortiz}
\author[2]{E. Perez}
\author[1]{\\D. Porquet}
\author[8]{B. Sicardy}
\author[1]{P. Vola}

\affil[1]{Aix Marseille Univ, CNRS, CNES, LAM, Marseille, France.}
\affil[2]{Instituto de Astrof\'isica de Andaluc\'ia (CSIC), Glorieta de la Astronom\'ia, E-18080 Granada, Spain.}
\affil[3]{AIM, CEA, CNRS, Universit\'e Paris-Saclay, Universit\'e Paris Diderot, Sorbonne Paris Cite, Observatoire de Paris, PSL University, F-91191 Gif-sur-Yvette Cedex, France.}
\affil[4]{Observatoire Astronomique de Strasbourg, Universit\'e de Strasbourg, CNRS, 11 Rue de l'Universit\'e, F-67000 Strasbourg, France.}
\affil[5]{DARK, Niels Bohr Institute, University of Copenhagen, Lyngbyvej 2, DK-2100 Copenhagen, Denmark.}
\affil[6]{Department of Physics and Astronomy, Stony Brook University, Stony Brook, NY 11794-3800, USA.}
\affil[7]{Universit\'e Cote d'Azur, Observatoire de la Cote d'Azur, CNRS, Laboratoire Lagrange, Laboratoire ARTEMIS, France.}
\affil[8]{LESIA/Observatoire de Paris, univ. PSL, CNRS, Paris, France.}

\begin{document}

\maketitle

\begin{abstract}
We present here the Calar Alto Schmidt-Lemaitre Telescope (CASTLE) concept, a technology demonstrator for curved detectors, that will be installed at the Calar Alto Observatory (Spain). This telescope has a wide field of view (2\fdeg36$\times$1\fdeg56) and a design, optimised to generate a Point Spread Function with very low level wings and reduced ghost features, which makes it considerably less susceptible to several systematic effects usually affecting similar systems. These characteristics are particularly suited to study the low surface brightness Universe. CASTLE will be able to reach surface brightness orders of magnitude fainter than the sky background level and observe the extremely extended and faint features around galaxies such as tidal features, stellar halos, intra-cluster light, etc.
CASTLE will also be used to search and detect astrophysical transients such as gamma ray bursts (GRB), gravitational wave optical counterparts, neutrino counterparts, etc. This will increase the number of precisely localized GRBs from 20\% to 60\% (in the case of Fermi/GMB GRBs).

\end{abstract}
\section{Introduction}
\begin{figure}
 \begin{center}
  \includegraphics[width=0.75\textwidth]{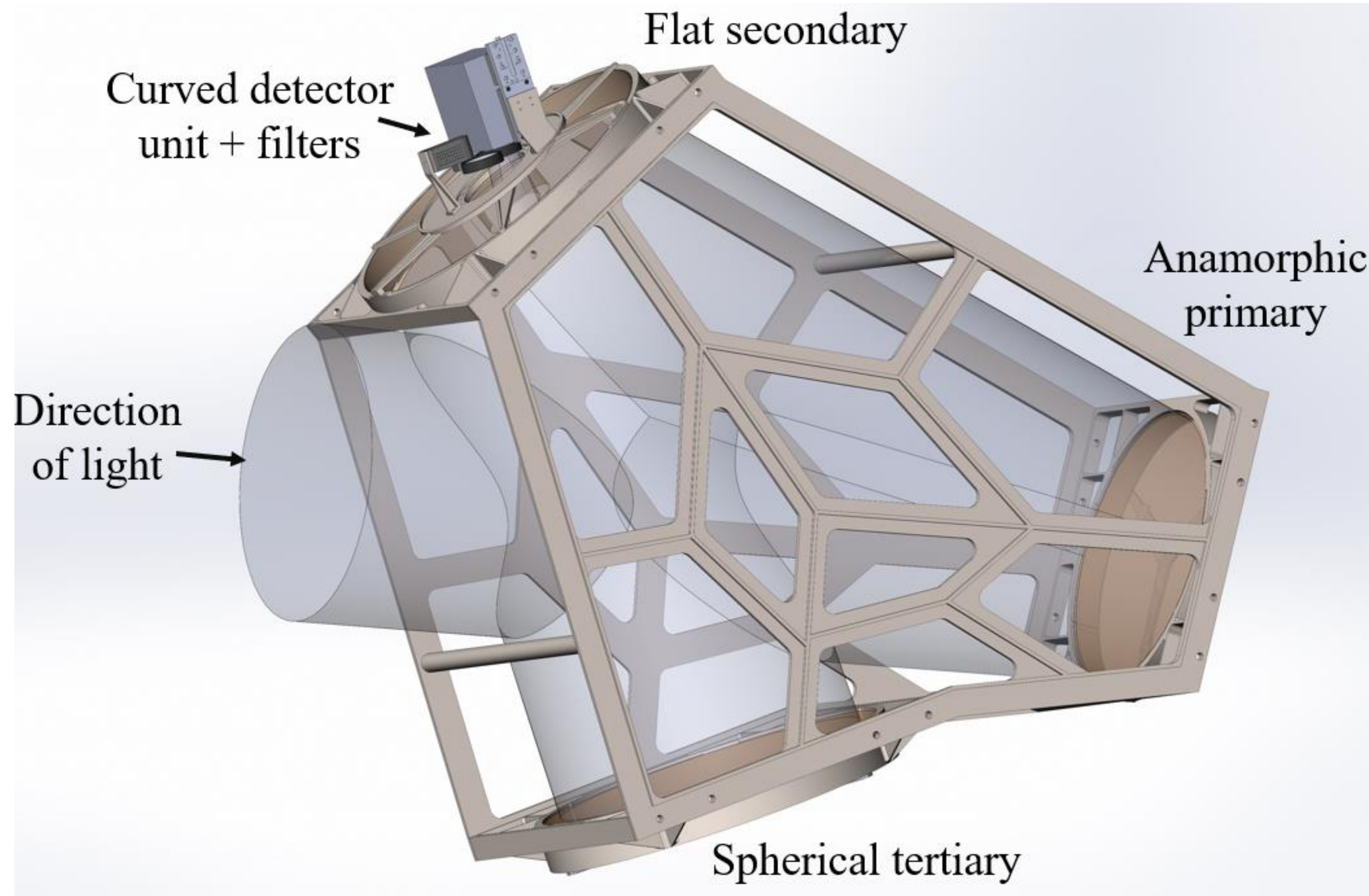}
  \caption{Opto-mechanical design of CASTLE: light reaches an anamorphic primary, is reflected by a flat secondary and
  a spherical tertiary yields a simple spherically-curved focal surface where a curved detector is placed \citep{Lombardo_2019c}.}
  \label{optical_design}
 \end{center}
\end{figure}

The Calar Alto Schmidt-Lemaitre Telescope (CASTLE) is a small sized telescope highly optimized for the detection of extremely faint and extended astrophysical objects. 
The telescope will be installed at the Calar Alto observatory (in Almeria, Spain).
CASTLE has an off-axis fully reflective Schmidt design \citep{lombardo_2019b, muslimov}, composed of an anamorphic primary mirror of $\sim$40 cm (whose purpose is to replace the entrance correcting Schmidt plate), a flat secondary mirror and a spherical tertiary mirror that focuses the light onto a convex focal surface (Figure~\ref{optical_design}). 

The advantages of such telescope come from its unique design that combines: a) the wide field of view of a Schmidt telescope (2\fdeg36$\times$1\fdeg56 with 1\arcsec per pixel), b) the removal of supporting spiders and related diffraction effects, c) the removal of field flattening optics provided by the usage of a curved sensor to match its curved focal surface. 
All these elements contribute to generate a Point Spread Function with very low level wings (Figure~\ref{psf_center}), uniform across the full field of view, and to effectively suppress ghost reflections, a capability that was verified by performing end-to-end photon Monte Carlo simulations \citep{lombardo_2019b} with the \texttt{PhoSim} software \citep{peterson_2015}.
The detector unit will have a science graded curved CMOS sensor (back-side illuminated to increase the Quantum Efficiency up to 80\%).
The telescope will be equipped with \textit{g}, \textit{r} and wide-band \textit{g+r} (luminance filter) filters for its science applications and the facility will be robotic.

CASTLE is proposed as technological demonstrator for the curved detector technology in the astronomical domain. 
Many optical systems, especially the ones that provide a wide field of view, generate a curved focal surface. 
Therefore, without a curved sensor that matches the shape of the focal surface, some lenses have to be placed before the sensor, in such a way that they project the image on its flat surface. 
This solution is not ideal as by doing so, one reduces the efficiency of the system (the lens does not transmit all the light despite the best AR coatings, all the while creating ghost reflections in the beam) and in general reduces the optical quality of the system. 
This is why the use of curved sensor can provide innovative solutions that allow the system to become more compact and less complex. 
In the recent years several prototypes of curved sensors (for commercial application) have been made and studied \citep{guenter2017,iwert_2012,dumas_2012,tekaya_2013,Itonaga_2014,gregory_2015, lombardo_2019a}, and due to all their advantages, they are now considered for several instruments for ground- and space-based science \citep{richard_2019}.

Considering its characteristics, CASTLE will be a tool for new discoveries in the domain of low surface brightness (LSB) and the transient search and detection.
The large field of view will be highly beneficial to search for gravitational wave optical counterparts, gamma ray bursts, neutrinos and fast radio bursts sources. 
The fast readout provided by its detector will also allow for the detection of transneptunian or trojan objects occultation in front of bright stars.

A LSB survey will be carried out during dark times, to limit the influence of bright sky background caused by the illumination from the Moon, and it will mostly focus on specific targets or limited areas on the sky. 
It will be possible, for example, to carry out a survey focused on observations of a sample of XUV galaxies during the first year of observations, then select specific tidal debris fields or ultra-diffuse galaxy sample on the second year while focusing part of the year on observation of the intra-cluster light in the Virgo cluster (visible during spring) or on the Ou4 outflow (visible in August).
More details on exposure times required for the observation and specific science cases achievable with CASTLE are provided in the following sections. 
Table~\ref{tab: survey} lists  current ground-based surveys used to observe the LSB Universe.
Such list does not intend to be exhaustive but only to show some typical characteristics and limiting surface brightness (SB) of large and mid/small sized telescopes. CASTLE will complement the current panorama.  

\begin{table}
\begin{center}
\caption{Example of surveys for LSB observations. Note that the SB has been measured with different metrics in these surveys (as described in their own references), which makes a direct comparison less immediate.}
\begin{threeparttable}
\begin{tabular}{lcccc}
\hline\\[-1.5ex]
Survey name & SDSS\tnote{1} & CFHT\tnote{2} & Burrell Schmidt\tnote{3} & Dragonfly\tnote{4} \\  
\hline\\[-1.5ex]
Primary diameter(m)& 2.5 & 3.6 & 0.9 & 0.143 (48 elements array) \\ 
Field of view  & 3\fdeg0 & 0\fdeg96$\times$0\fdeg94 & 1\fdeg65$\times$1\fdeg65 & 2\fdeg6$\times$1\fdeg9 \\ 
Pixel size  & 0\farcsec396 & 0\farcsec197 & 1\farcsec45 &  2\farcsec8 \\ 
SB(\sbmag)  & 26.4 & 28.5 & 28.5 & 29.8 \\ 
Filter & \textit{g} & \textit{g} & \textit{V} & \textit{g} \\ 
\hline\\[-1.1ex]
\end{tabular}
\label{tab: survey}
\begin{tablenotes}\footnotesize
\item[1] \citet{kniazev_2004}, $^2$ \citet{duc_2015}, $^3$ \citet{mihos_2017}, $^4$ \citet{danieli_2020}.
\end{tablenotes}
\end{threeparttable}
\end{center}
\end{table}
\section{Main performances}
\begin{figure}
 \begin{center}
  \includegraphics[width=0.75\textwidth]{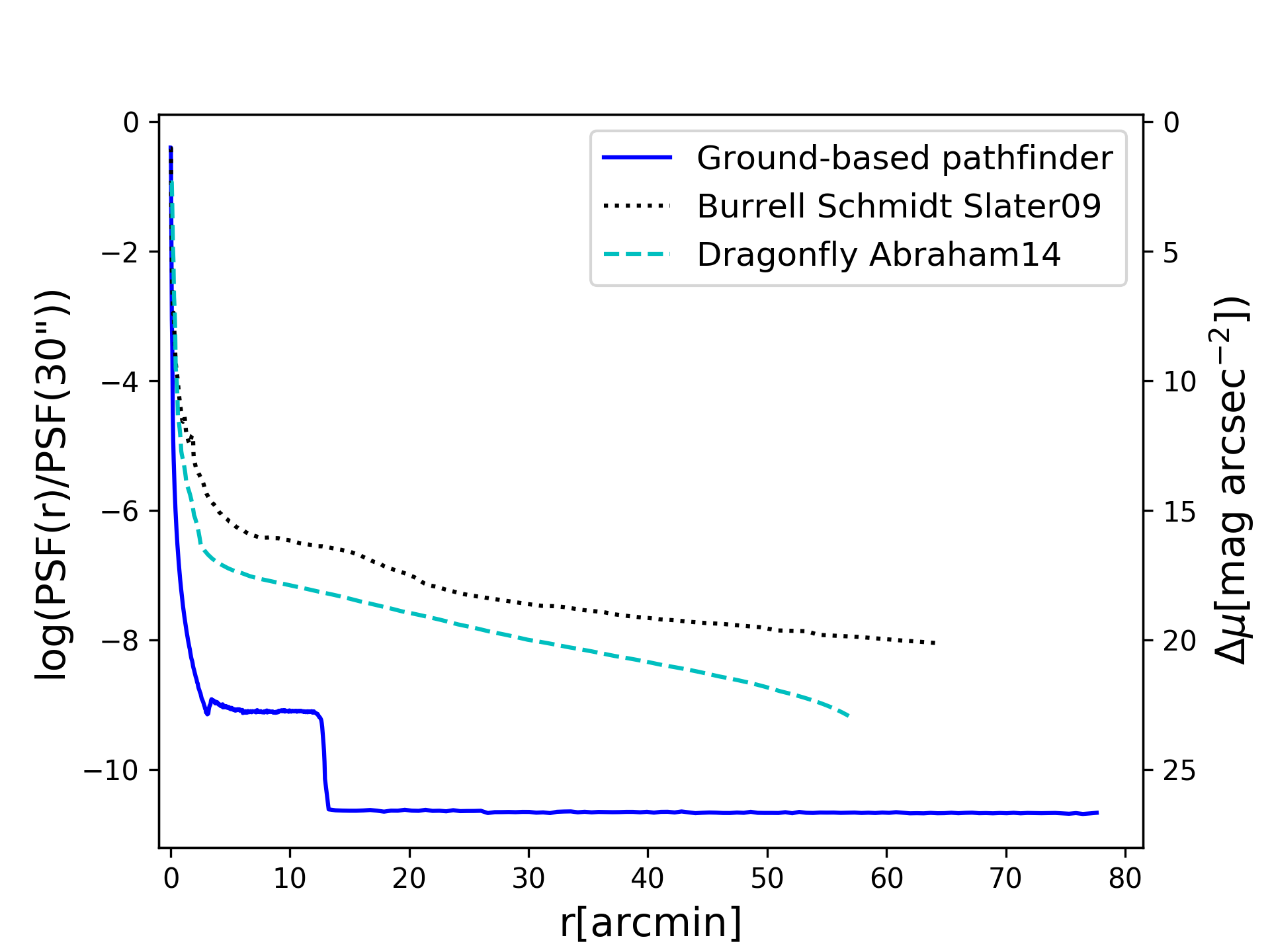}
  \caption{PSF of CASTLE from photo Monte Carlo simulations in solid blue line, PSF of Dragonfly in cyan dashed line  and PSF of the Burrell Schmidt in dotted black line. All curves are normalized with their values within 30\arcsec ~\citep{lombardo_2019b}. }
  \label{psf_center}
 \end{center}
\end{figure}

Extensive end-to-end photon Monte Carlo simulations were performed to ascertain the performances of the telescope and the results showed that the telescope PSF reaches very low wing levels and also limits  ghost reflections (Figure~\ref{psf_center}). 
They also demonstrated the overall uniformity of the PSF shape across the full field of view where the differences of the PSF at large scales are negligible. 
The corresponding FWHM under observing seeing conditions of $\sim$1\arcsec ~resulted to be $\sim$2\farcsec7, which will be properly sampled with CASTLE's pixel scale of 1\arcsec.

\begin{figure}
 \begin{center}
  \includegraphics[width=1.\textwidth]{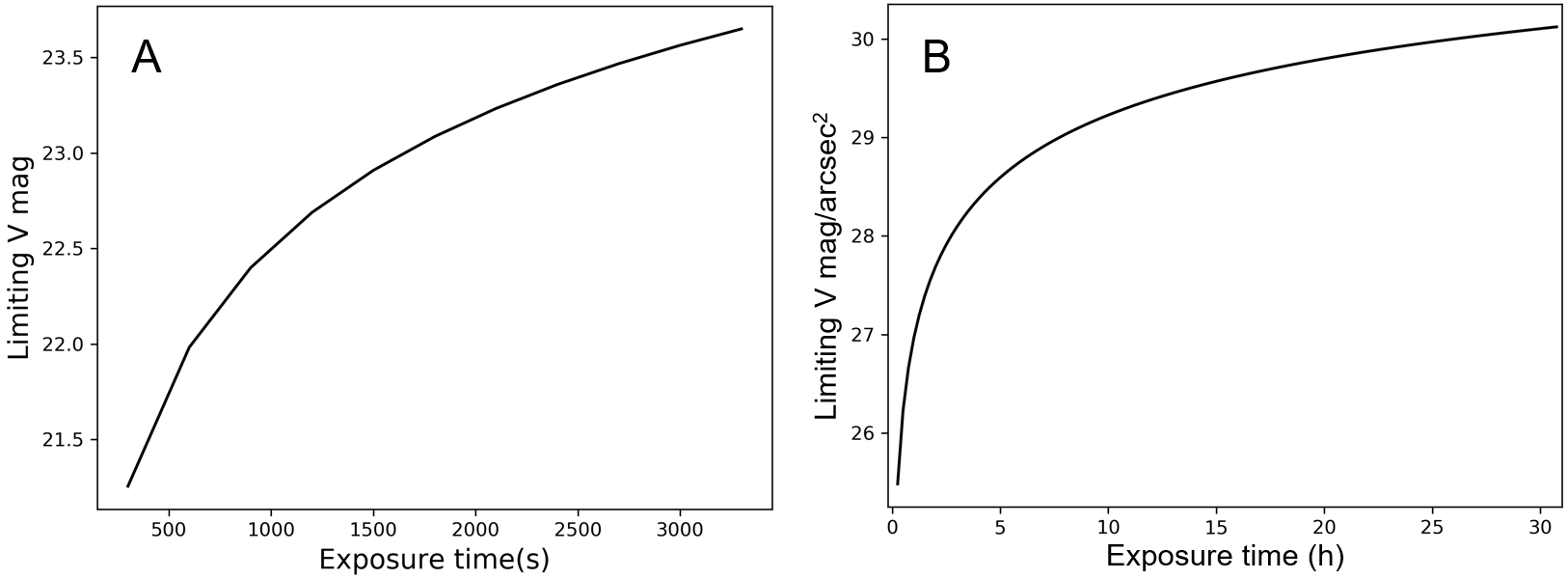}
  \caption{Limiting magnitudes for CASTLE (SNR=5) in \textit{V} band as function of exposure time. A: limiting magnitudes for point source observations made by stacking frames of 300 s exposure time each. B: limiting surface brightness for extended sources made by stacking frames of 900\,s exposure time each.}
  \label{lim_mag}
 \end{center}
\end{figure}

CASTLE will be used for various science cases (e.g., star like and extended sources) reaching the limiting magnitudes and limiting surface brightness magnitudes listed in Figure~\ref{lim_mag}. 
For these we considered the moonless night-sky surface brightness \citep{sky_caha} in the \textit{V} band, \textit{V}~=~22.01~\sbmag, and the other values listed in Table~\ref{tab: data}. 
All the values for the site conditions listed in Table~\ref{tab: data} are from \cite{sky_caha}. The seeing value of 1\arcsec\, used in the simulations is higher than the median seeing of 0\farcsec9 recorded for the CAHA site over a year \citep{sky_caha}.

\begin{table}
\begin{center}
\caption{Parameters of telescope and observing site.}
\begin{tabular}{lr}
\hline\\[-1.5ex]
Telescope parameter & Value\\
\hline\\[-1.8ex]
Field of view  &   2\fdeg36$\times$1\fdeg56\\ [0.25ex]
F/\#  & 2.5\\ [0.25ex]
Diameter  & 356 mm \\ [0.25ex]
Detector shape/radius of curvature & Convex/$\sim$800 mm\\[0.25ex]
Total throughput & \begin{tabular}{c}
$>40$\%\,${\lambda}\in $[400, 600] nm\\%[-0.5em]
$>50$\%\,${\lambda}\in$ [600, 1000] nm\end{tabular}\\ %[0.25ex]
\hline\\[-1.5ex]
Detector parameter & Value\\
\hline\\[-1.8ex]
Pixel size on sky & 1\arcsec\\[0.25ex]
Quantum Efficiency & >80\% @520 nm\\[0.25ex]
RON & <2 e$^-$\\[0.25ex]
fps & >100\\[0.25ex]
\hline\\[-1.5ex]
Site parameter  & Value\\
\hline\\[-1.8ex]
Seeing & 1\arcsec \\[0.25ex]
Sky background V band & 22.01 mag arcsec$^{-2}$\\[0.25ex]
Clear hours for observations/year & 70\%\\[0.25ex]
Photometric nights/year & 30\%\\[0.25ex]
\hline\\[-1.1ex]
\end{tabular}
\label{tab: data}
\end{center}
\end{table}

The point source limiting magnitude is computed by considering exposures integrated over 300 s, and then averaged over the exposure times shown, up to a 1h integration. 
This provides an example of observing strategy that allows to sample the time variability of the light emitted by the targets, while keeping the exposure time high enough to observe also fainter objects.
In the case of the surface brightness, instead, the integration is performed over exposures of 900 s each, keeping the contribution of RON to the global noise budget low (sky dominated signal).
For both cases we extract the flux on an aperture of 3$\sigma$ corresponding to a radius on sky of 6\arcsec. 
Given the PSF effective dimension, we expect the limiting surface brightness to be affected in some amount by star crowding. 
This is usually overcome by using complementary images with better image quality (e.g. from CFHT). 

\section{The low surface brightness Universe}

The observation of ultra-low surface brightness Universe still remains a largely unexplored area of astrophysics  outside the Local Group. This is mainly due to the extremely low brightness features (>29\sbmag), magnitudes fainter than the sky background, with a physical extension  on the sky that reach from a few arcminutes to many degrees.
Even though several objects and features have been recently detected \citep{martinez_delgado_2010,dokkum_2015,knapen_2017,mihos_2017,abraham_2014,duc_2015,boissier_2008}, many more need to be studied, since their observations require optimized systems and long exposure times. 
Even when observing with larger surveys, the requested time, though being much shorter, might be assigned over large time intervals of a few years.
Additionally, by providing redundant observations, we can increase the number of independent systems and surveys to verify the published results (when required).

Due to its extremely low level PSF wings and large field of view, CASTLE will be able to observe a vast typology of this class of objects and features, and it will significantly contribute to new discoveries and advancement in this field. 
In particular the LSB universe can be used to test the $\Lambda$CDM cosmology model by, e.g., observing the abundance, location, and properties of tidal features and dwarf satellite galaxies (Section~\ref{sec:tidal_tailes}),  giant low surface brightness galaxies (Section~\ref{sec:glsb}), diffuse intra-cluster light (Section~\ref{sec:icl}) and ultra-diffuse galaxies (Section~\ref{sec:udg}), that are all believed to be pieces of the puzzle of the hierarchical model for galaxy formation and evolution, and are still under-sampled in current catalogues due to their low surface brightness and large extension on the sky.
It is important to notice the role of the HI component, found for several of these features, that provides complementary information to the visible band and enrich our understanding of the topic (Section~\ref{sec:HI}).

LSB features are not only indicative of interactions between galaxies, but they can also come from outflows whose origin is still under scrutiny, as for the Ou4 giant bipolar outflow \citep{acker12, corradi14}. In this case the ouflow is so extended and faint that it requires the same efforts outlined for LSB observations (Section~\ref{sec:ou4}).
In the following subsections all these science cases are explained in greater detail.

\subsection{Stellar streams and tidal debris from galaxies interactions}
\label{sec:tidal_tailes}

\begin{figure}[htb]
 \begin{center}
  \includegraphics[width=0.75\textwidth]{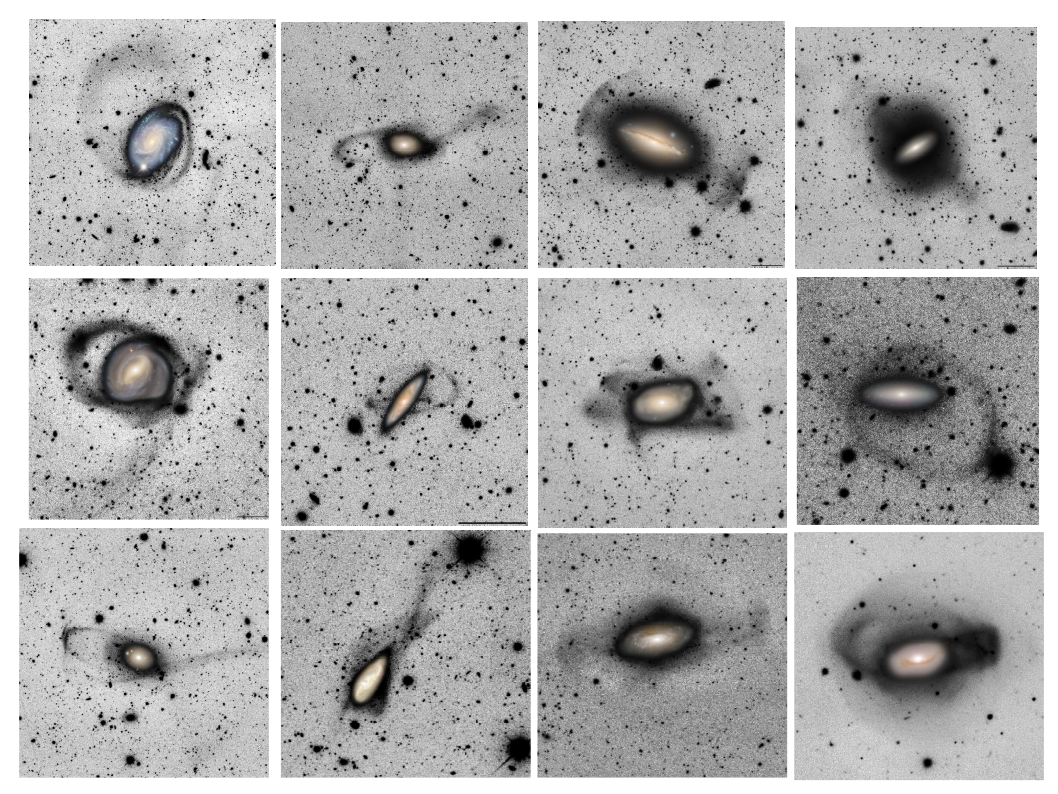}
  \caption{Stellar tidal streams around nearby galaxies discovered in the deep images from the Legacy surveys (Mart\' \i nez-Delgado et al. 2020)}
  \label{fig:streams}
 \end{center}
\end{figure}

\begin{figure}[htb]
 \begin{center}
  \includegraphics[width=0.50\textwidth]{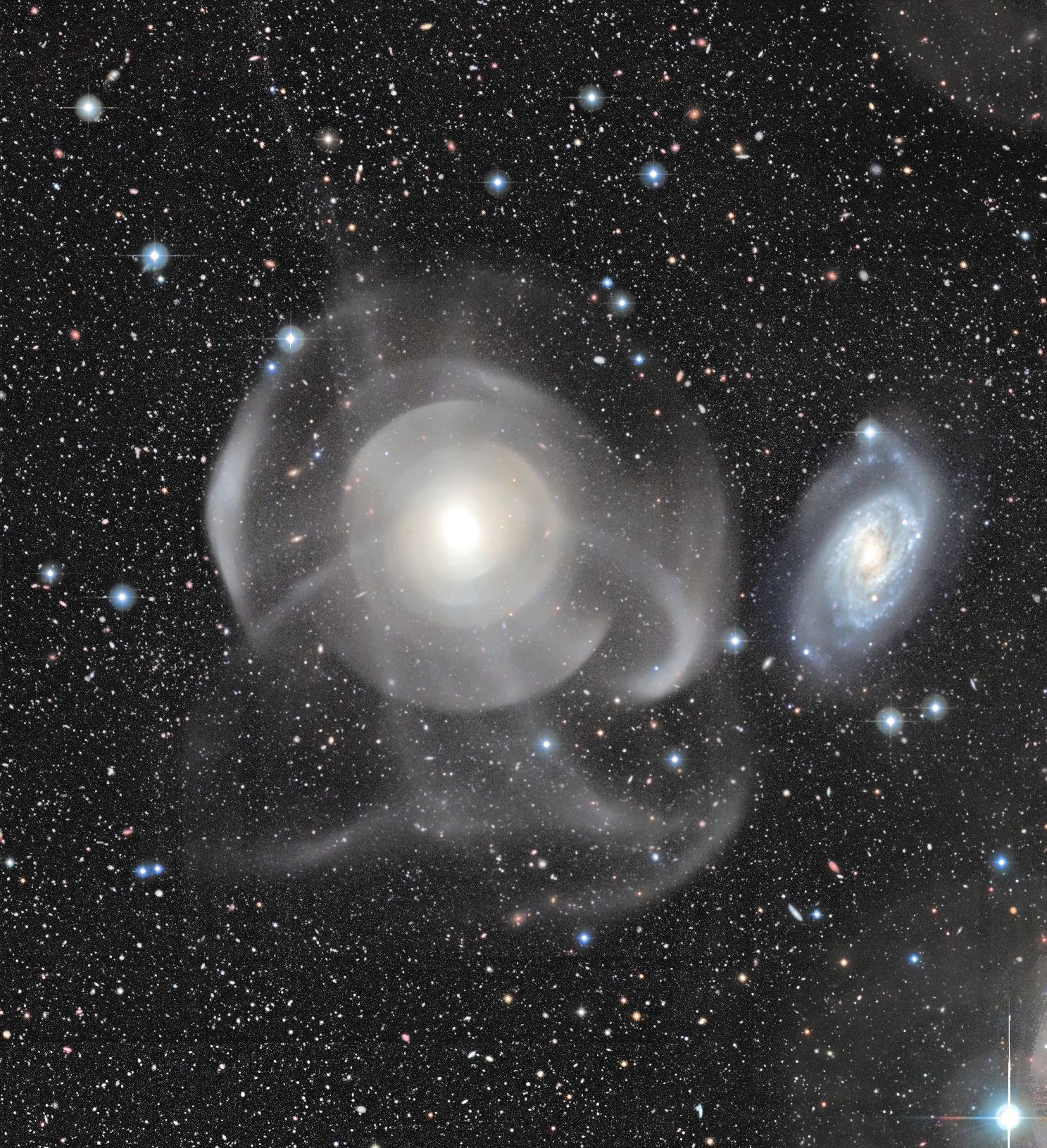}
  \caption{Shells and tidal tails around the Early-Type galaxies NGC 474. Image credit: CFHT/Coelum - J.-C. Cuillandre \& G. Anselmi}
  \label{NGC0474}
 \end{center}
\end{figure}

The $\Lambda$CDM cosmology predicts a hierarchical assembly of galaxies through a series of mergers of smaller galaxies in addition to continuous accretion of gas and dark matter \citep{white_1978, bullock_2005, naab_2007, cooper2013, cooper_2015, rodriguez_gomez_2016}.
The tidal interactions during these mergers leave highly deformed remnants around the galaxies (Figure~\ref{fig:streams}-\ref{NGC0474}).
These tidal features, such as stellar streams, shells, tails and bridges, remain around the galaxies for a long time
as their dynamical,  diffusion and survival timescales are about a Gyr or greater, depending on the types of structures \citep{mancillas_2019}.
Therefore, they are important clues to understanding the mass assembly history of galaxies. 

Many tidal features are discovered and resolved into individual stars in Local Group galaxies, revealing the imprints of past mergers.
The Milky Way contains multiple coherent stellar structures associated with tidally disrupted satellites \citep{grillmair_2006, belokurov_2006, juric_2008}.
An archetype is the Sagittarius stream \citep{ibata_1994, yanny_2000,Majewski_2003} and, similarly, the Andromeda galaxy hosts several stellar streams \citep{ibata_2001, ibata_2007, ferguson_2002, richardon_2008, McConnachie_2009,gilbert_2012}.
These are signatures of ongoing minor mergers, as well as a stellar bridge connecting Andromeda to Triangulum, likely due to a recent major interaction \citep{McConnachie_2009}.

The detection of these faint tidal remnants in nearby galaxies is a ubiquitous aspect of galaxy formation that has not yet been fully exploited, mainly because they are challenging to observe. 
Individual stars cannot be entirely resolved beyond the Local Group, and the tidal features appear as extended, low surface brightness (LSB) structures. Over the last decade, the {\it Stellar Tidal Stream Survey} \citep{martinez_delgado_2010,martinez_delgado_2019} has revealed for the first time an assortment of large-scale tidal structures in their halos by means of deep images of nearby spiral galaxies taken with robotic amateur telescopes. A more systematic search for tidal streams has been recently carried out in deep images from large-scale photometric surveys (e.g. SDSS: \cite{morales_2018} and the Legacy surveys: Mart\'\i nez- Delgado et al. 2020; see Figure~\ref{fig:streams}). These observations have yielded so far an unprecedented sample of a $\sim$ few hundreds of bright stellar streams in nearby spiral galaxies, including the discovery of observational analogues to the canonical morphologies found in cosmological simulations of stellar halos. 

This sample of stellar streams offers a unique opportunity to study in detail the apparently still dramatic last stages of galaxy assembly in the local Universe and to probe the anticipated estimates of frequency of tidal stellar features from the LCDM paradigm for Milky Way-sized galaxies. However, the observations of these LSB structures have been difficult because of the presence, even after flat-fielding, of remaining extended background fluctuations due, among others, to scattering by atmospheric aerosols and dust, and reflections and scattering between the telescope and camera  optical elements \citep{mihos05, duc_2015,sandin_2014}. 
Significant improvements have been made in such observations, by optimizing observational strategies \citep{trujillo_2016,duc_2015} or instruments \citep{abraham_2014, van_Dokkum_2014}. 
These internal reflections of light from bright objects can be particularly difficult to spot when coming from bright galactic nuclei, since they can be mistaken for real stellar halos \citep{sandin_2014, michard_2002, duc_2015}.
Since they can be more dominant in certain colors, their missed identification, e.g. when they are smaller than the apparent size of the galaxy, can potentially lead to a misinterpretation of the stellar population of the outer stellar halo of the galaxy \citep{karabal_2017}.
Having a telescope with minimized ghosts and scattering light effects can significantly improve our knowledge of  outer stellar halos, their shape and profiles (breaks, truncation, etc.).

CASTLE can make a significant contribution to the studies of tidal features and stellar halos around galaxies and reveal the history of galaxy formation.
The field of view of individual images is large enough to cover an area as far as the virial radius of MW-like galaxies beyond the distance of $\sim 11$ Mpc.
Therefore, many nearby galaxies are good targets for the search of surrounding tidal debris within their gravitational influences.
Additionally the $\Lambda$CDM model shows a shortage of dwarf satellite galaxies around normal galaxies (there are less observed than what is predicted by the model), and some of those dwarfs might have been shredded apart by the tidal field of the parent galaxies.
Observations with CASTLE can simultaneously find diffuse satellite galaxies as well as tidal debris, and thus, they can put an important observational constraint on the $\Lambda$CDM model.

The tidal debris field is an important laboratory for studying not only the destruction, but also the formation of stellar structures around galaxies.
Tidal interactions between galaxies can strip and expel the material from the interacting galaxies.
The tidal force impacts the galaxies' outskirts more significantly than the inner parts, which often produce huge tidal tails of HI gas as the gas is typically more extended than stars around galaxies.
These tidal tails, if they are dense, may fragment and form dwarf galaxies, namely tidal dwarf galaxies \citep{Duc_2013}.
In fact, the GALEX satellite found tantalizing evidence for star formation in the ourskirt of and outside nearby galaxies \citep{neff_2005}.
The UV emission detected by GALEX is from O or B-type stars, some of which appear to be associated with tidal tails, and indicate the rate of recent star formation.
However, it remains unknown as to how many stars have been formed and built up in these environments in the past.
CASTLE can detect the optical features of low surface brightness associated with the UV emission and show the build-up history of the galaxy's immediate environment.

Historically, HI tails were not always found accompanied by optical counterparts. This has changed in recent years, and more optical counterparts to the HI tails are being found as the limiting depth increases \citep{mihos05,martinez_delgado_2010, ferrarese_2012, muller_2019a}. On the other hand there may be cases where an HI filament truly has no optical counterpart, like the Magellanic stream. Besides, gaseous tail tend to evaporate with a quicker timescale than their optical couterparts; therefore the census of diffuse optical tails may reveal older mergers. 
CASTLE will show where, when and how many stars could form in the tidal field around galaxies.
By observing both the destruction and formation, we can understand the eco-cycle in the vicinity of galaxies.

\subsection{Galaxies with extremely extended and faint disks}
\label{sec:glsb}

We described before the evidence for star formation in the ourskirt of and outside nearby galaxies in the form of UV emitting tidal tails (Section~\ref{sec:tidal_tailes}), found by GALEX.
This UV satellite has also discovered eXtended UV (XUV) emission around otherwise normal nearby galaxies \citep{gidepaz05,thilker05,thilker07}. This was possible because of its large field of view
(about 1\deg\ diameter) and its deep images owing to the low UV sky background. This emission has been interpreted as the sign of wide-spread star formation. Understanding the conditions of this distant star formation is of paramount importance as it is related to how galaxies grow. For instance \citet{koda12} try to put constraints on the IMF in the outer disk from \textit{u} and H$\alpha$ images.

Independently, so called ``giant low surface brightness galaxies'' (GLSBs) had been found in deep optical images since the 1980s. They are galaxies with low surface brightness (extrapolated central surface brightness of a faint disk) but very extended disks, in which star formation is probably occurring, as in the more extreme example, Malin 1 \citep{boissier16,junais20}. 
Up to now, their study has been difficult but the technical progresses has allowed astronomers to revisit such objects in the recent years, and even discover new ones \citep{hagen16}. Most GLSBs host in their center a denser region that may be considered as a ``normal'' galaxy. 
So that XUV galaxies and GLSBs could be considered as the facets of similar physics (extended disks around otherwise relatively normal galaxies in which star formation occurs).
One way to test this hypothesis is by constraining the optical properties of the XUV detected disks, in order to compare them to the standard definition of GLSBs \citep{sprayberry95}.

\begin{figure*}[htb]
\centering
\includegraphics[scale=0.315, angle=0.0]{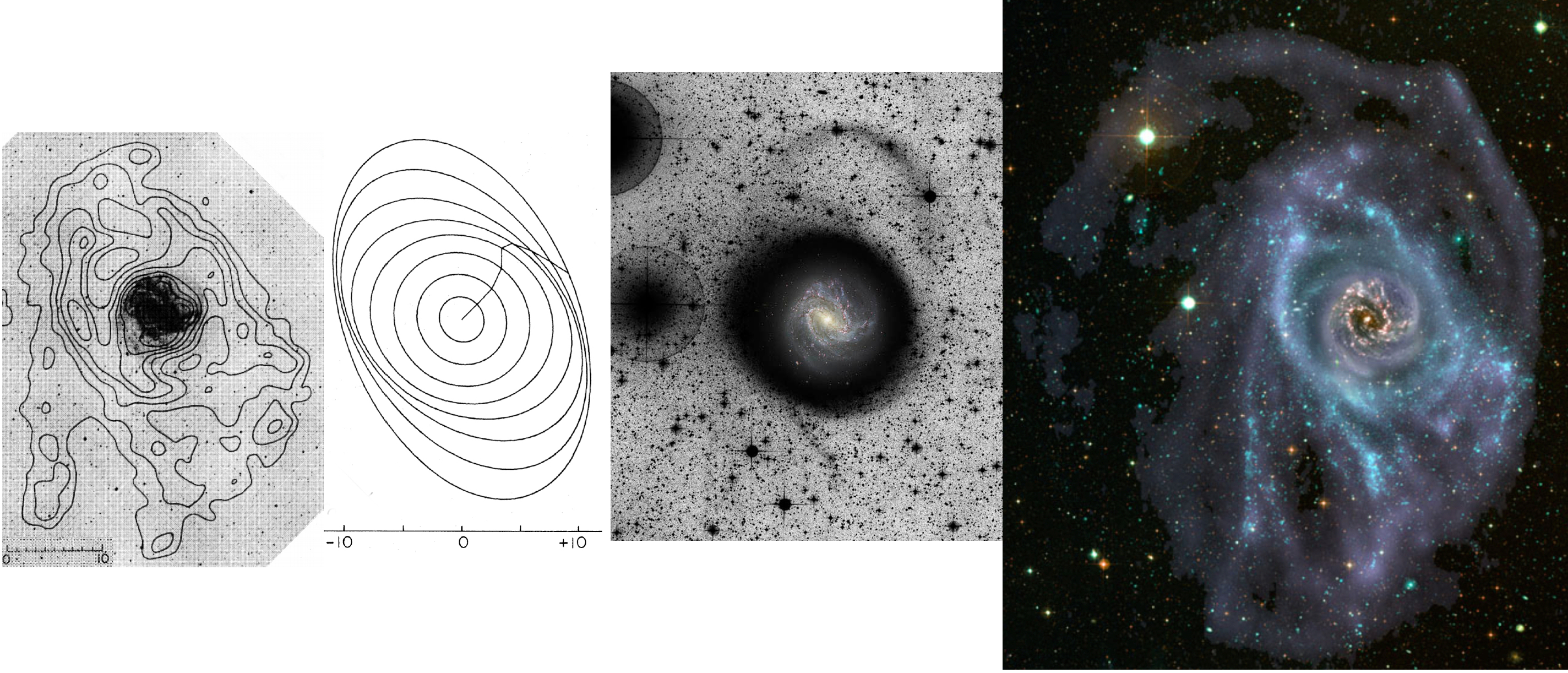}
\caption{{\it Left} two panels: H{\sc i} distribution and tilted ring model of M83 (reproduced with permission
from \citealt{rogstad_74}); next to it a deep optical image obtained by \citet{malin_97} with a shallower colour image superimposed (credit: D. Malin). The {\it rightmost} image was
constructed from {\it GALEX} UV data, optical images, and the H{\sc i} distribution obtained in the Local Volume HI Survey conducted with CSIRO's Australia Telescope Compact Array (ATCA; credit: B.~Koribalski and A.~R.~L\'opez-S\'anchez). All images are on the same scale, and the units in the left two panels are in arcminutes.}
%\index{M83}
\label{m83}
\end{figure*}

\citet{thilker07} has identified tens of XUV galaxies. 
More recently, Bouquin et al. (in preparation, also presented in IAU355 symposium in Tenerife, 2019) found 349 XUV galaxies out of a total of 1931 galaxies in the S4G sample with optical diameter larger than 1\arcmin, and for which GALEX data were
available.
The most spectacular are nearby galaxies such as M83 (Figure~\ref{m83}), that to be fully covered with GALEX needed mosaics of several pointings (and may need a few pointings with CASTLE too). 
Other XUV galaxies are at different distances offering us a range of sizes. 
The D25 distribution in the XUVs of \citet{thilker07} and \citet{bouquin_2018} actually peaks around 2\arcmin (keeping in mind the UV emission extends up to 2 to 4 times this size in many galaxies) with 13 XUV galaxies in the S4G sample above %10\arcsec. ?? [Albert]
10\arcmin.

%Albert : added figure

With its FOV of 2\fdeg36$\times$1\fdeg56, CASTLE  will thus be an efficient instrument to characterize the optical properties of all extended disks found in the UV with GALEX.
Since the galaxies are larger than 1\arcmin, the spatial resolution of the arcsec range is largely sufficient to e.g. measure scale-lengths, and the large FOV is similar to the GALEX one, thus allowing us to detect structures on similar scales. 

Observations of non XUV galaxies with matching properties will provide a control sample to check if the presence of XUV disks and GLSBs are really related or not. 
The data from such samples will also be useful to study other aspects linked to the environment of galaxies in general, owing to the large FOV. 
For instance, most of the XUV galaxies in the  \citet{thilker07} sample are beyond a distance of 17 Mpc that would correspond to a virial diameter of 300 kpc at one degree, making these galaxies good targets for satellite searches.

\subsection{The diffuse intra-cluster light}
\label{sec:icl}

In a Universe where the hierarchical formation and evolution is the driving mechanism in determining the current epoch characteristics of galaxies, it is expected that an abundance of low surface brightness intra-cluster tidal debris from disrupted systems, and an ubiquity of diffuse stellar component permeate the intra-cluster medium of galaxy clusters. In fact, in galaxy clusters a fraction of their baryonic content is represented by the intra-cluster light (ICL, or ``diffuse light''), a component that is gravitationally unbound to cluster galaxies, but bound to the cluster potential \citep[e.g.,][]{murante04,rudick10,dolag10,contini14,longobardi15,longobardi18}. Galaxy interactions, as well as tidal interactions between galaxies and the cluster potential, are both believed to play an important role in the production of the ICL. Hence, the ICL is thought to be intimately linked to the dynamical history of the cluster, so that the ICL's observable features
contain information about the evolutionary processes that took place in these dense environments. Moreover, it has been shown that ICL closely follows the global dark matter distribution of the hosting cluster, providing an accurate luminous tracer of dark matter \citep{montes19,asensio20}.\\
Despite its dynamical definition and its low surface brightness, the ICL is usually identified on the basis of its photometric properties. It is either identified as any optical light below a fixed surface brightness limit \citep{feldmeier04,mihos05,zibetti05,mihos_2017}, or as the less concentrated light profile that overlaps onto the one of the galaxy's halo \citep{gonzalez05,seigar07}. These studies have shown that the ICL can be found in the form of discrete tidal features, or as a diffuse halo around galaxies. In the latter case it is a discernibly separate entity from the host galaxies, with well defined transitions in the surface brightness profile, axis ratio, and position angle, and whose evolution is tied to the cluster as whole rather than to single cluster galaxies.

CASTLE, supplied of two $g$ and $r$ photometric passbands, will deliver accurate surface photometry and $g-r$ colours at competitive surface brightness limits. Its sensitivity, together with its large field of view and its design optimised to generate optimal PSF and reduced ghost features, will make it a powerfull machine to conduct systematic wide-field mapping of the low-surface brightness diffuse light component in the local and nearby Universe out to a distance of $\sim$ 20 Mpc. It will make it possible, for example, to entirely map the Virgo cluster, the nearest largest concentration of mass. At a distance of 16.5 Mpc, Virgo extends over an area of $\sim 104 $ deg$^2$. Here, the ICL is measured with surveys that reach a 3 $\sigma$ limit in the V-band of $\mu_{\mathrm{V,lim}}\sim 28.5\, \mathrm{mag\, arcsec^{-2}}$.  CASTLE can reach the same sensitivities in $\lesssim 10$ hrs, allowing us to complete the mapping of the entire cluster in $\sim 370$ hrs (we have included $\sim$10 deg$^{2}$ overlap between adjacent fields to ensure photometric uniformity across the survey.) 

Thus CASTLE will make it possible to produce the data with which to compare and test cosmological models.

\subsection{Ultra Diffuse Galaxies in and outside clusters}
\label{sec:udg}

Despite the long-standing knowledge of the existence of low surface brightness galaxies \citep{sandage_1984,bothun_1987,davies_1988,davies_1989,impey_1988,mcgaugh_1994,dalcanton_1997,oneil_1997,oneil_1999,binggeli_1998}, their significance in the full galaxy population has remained unrecognized.
The situation has been drastically altered by the recent discovery of ultra-diffuse galaxies (UDGs) \citep{dokkum_2015,koda_2015}.
The UDGs are as large
as giant galaxies in size, with the effective radii of $r_{\rm e}\sim 1$-$5$ kpc, but as faint as dwarf galaxies in luminosity, the equivalent stellar masses of only $\sim10^{7-8}{\,\rm M_{\odot}}$.
The most important is their abundance - UDGs are as abundant as
$L_{\star}$ galaxies in galaxy clusters \citep{koda_2015, vanderBurg_2016,vanderBurg_2017}.
This population of extreme low surface brightness galaxies may provide important clues to modeling galaxy formation and evolution \citep{yozin_2015,di_cintio_2017,Baushev_2018,ogiya_2018,carleton_2019,jiang_2019,liao_2019,Freundlich_2020}.
CASTLE can contribute significantly to elucidating the nature and origin of the UDGs.

Broadly, two hypotheses have been proposed for the origin of UDGs and are still debated.
The ``failed $L_{\star}$ galaxy'' hypothesis assumes that UDGs reside in massive dark halos,
but somehow failed to produce as many stars as typical $L_{\star}$ galaxies \citep{dokkum_2015}.
An alternative model, namely the ``puffed-up dwarf'' hypothesis,
suggests that the majority of UDGs are an extension of dwarf galaxy populations,
whose gas and stellar components are puffed up by weakened gravity due to ram pressure stripping,
tidal encounters, or feedback \citep{yozin_2015,di_cintio_2017, Baushev_2018}.
The baryons could also be intrinsically extended due to large initial spins \citep{amorisco_2016}.
Each of these physical processes leaves its own imprints on the photometric properties of UDGs,
their spatial variations within and outside galaxy clusters, and their relations to dwarf galaxies.
Thus, it is important to conduct a census of UDGs and their properties
in relation to giant and dwarf galaxies and as a function of their environments.
Image artifacts due to telescope optics have been the hindrance
to connecting the bright and faint galaxy populations, as they could not be detected and analyzed simultaneously in a single set of observations in an unbiased manner.
The superb image quality enabled by CASTLE will permit us to analyze the spectrum of galaxies from giants to dwarfs, including UDGs, so that their photometric properties can be derived and analyzed in an unbiased way.  

Numerous studies have found UDGs in many clusters and groups,
in large-scale structures, and in the field by now \citep{mihos_2015,munoz_2015,makarov_2015,vanderBurg_2016,vanderBurg_2017,merritt_2016,martinez_delgado_2016,roman_2017,shi_2017,greco_2018}.
Small telescopes have played key roles in finding UDGs, by utilizing
large photometric apertures ($\gtrsim4$-$6\arcsec$) suitable to the large angular extents of UDGs \citep{abraham_2014,dokkum_2015, mihos_2015,mihos_2017,shi_2017}.
However, most studies suffer from significant biases,
due to the image artifacts introduced by complex optical structures \citep{koda_2015,yagi_2016}
and to source confusion \citep{dokkum_2015}.
The simple optics of CASTLE can make a breakthrough.

The abundant presence of UDGs may cause a potential conflict with the cosmological model.
Thus, studies of this new galaxy population are statistical in nature.
Follow-up spectroscopy is severely limited to few case studies \citep{van_dokkum_2016,van_dokkum_2019}.
A viable next step, in conjunction with spectroscopic case studies, is a systematic imaging survey,
providing a reliable census of the UDG population
in relation to giants and dwarfs.
CASTLE is an ideal instrument for such a complete census.
We will confirm whether UDGs are distinct from, or a continuation of, dwarf and/or giant galaxies, by analyzing their internal and statistical properties. We will constrain/identify physical mechanisms for the origin of UDGs, by analyzing their environmental dependence and spatial variations, within and beyond the virial radii of the local poor-to-rich clusters. For example, we will correlate the properties of UDGs (e.g., tidal features, blue color, large
axis ratio, and orientation) with their environments.

\subsection{Synergies with HI surveys}
\label{sec:HI}

The HI has been extensively used to gather morphological and kinematics information on the cold gas in and around galaxies. These data further constrain gas accretion and removal processes in galaxies defining, in this way, their gas reservoir and consequently the build up of stellar mass. 

\begin{figure*}
\centering
\includegraphics[scale=0.367, angle=0.0]{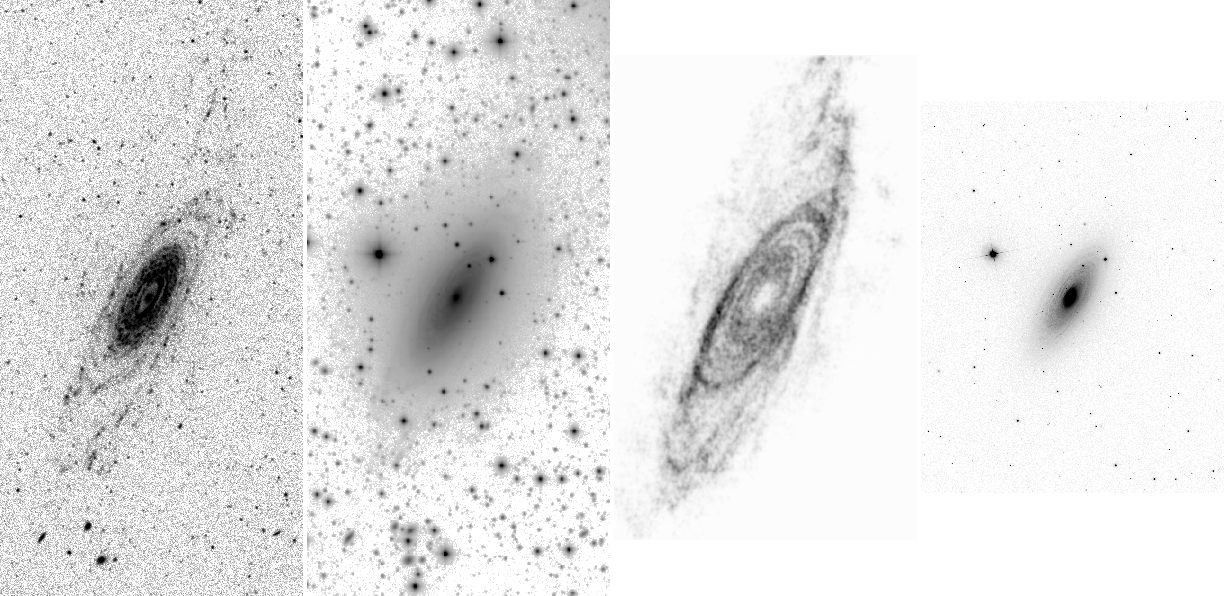}
\caption{Comparison of images of NGC 2841. At left:
GALEX far UV image; next to it the Dragonfly (\citealt{abraham_2014}) G-band image (\citealt{zhang_18});
then the integral HI image from the THINGS survey (\citealt{walter_08}); and at right the SDSS r-band image.}

\label{N2841}
\end{figure*}

We have already seen the importance and complementary of information brought by the HI component in several of the science cases examined earlier. 
For instance we saw that the HI component is found in tidal features and outskirt of galaxies (Section~\ref{sec:tidal_tailes}) hinting at recent star formation activities in those regions.
Many of the XUV galaxies (Section~\ref{sec:glsb}) could also be linked to large HI disks, and the relation between faint stellar populations and diffuse gas is an open issue. One of the most intriguing Dragonfly observations concerns the outskirts of the galaxy NGC 2841, which is 
known to have an extended and warped HI envelope, and a type I extended UV disc. \citet{zhang_18} 
detect faint optical emission in this envelope, and claim the existence of a faint stellar disc, with a stellar to gas mass ratio of about 3:1. 
We show a composite image of this case in Figure~\ref{N2841}. Such observations can be repeated with CASTLE, with the aim of producing deeper images thanks to its well behaved PSF (cf. Figure~\ref{psf_center}), and extended to numerous similar cases. 

Deep optical imaging has been prompted to answer questions of whether the extended HI was
associated with faint optical emission. 
However, some of the answers were not as convincing, as evidenced by the discrepancy in the observation of the faint emission around NGC 5907 \citep{martinez_d08, van_dokkum_19, muller_2019b}. 
Deep surveys of these objects with CASTLE will shed more light on this problem.

The HI in nearby spiral galaxies has been studied for about 50 years or more, with both single dish radio telescopes as well as interferometers. Most of the observations have been pointed towards pre-selected 
objects, whose morphology and rough extent were already listed in catalogues. The extent of the HI is 
variable, and quite a number of them extend to 3  - 5 times the optical isophotal radius, as measured at
the 25th \sbmag isophote in the blue (e.g. Figure 17 in \citealt{bosma_17}). Such galaxies, if they are regular, are ideal for studying their mass distribution from the rotation curves (e.g. \citealt{bosma_78}), which
remain flat well beyond the optical image, thereby indicating the presence of dark matter. A modern 
compilation resulted in the SPARC sample of 175 HI rotation curves, and has been extensively used for 
discussion of the dark matter problem (\citealt{lelli_16}, and following).

Instrumental developments have lead to the construction of ``phased array feeds'', which can equip the
focal plane of individual radio telescopes in an interferometer, thereby enlarging the field of view (see
\citealt{verheijen_08} for an early test result). This instrumental approach is driving the efforts of the radio
astronomical community in the construction of the Square Kilometer Array (SKA). At present, a 
SKA-precursor, the ASKAP telescope, is in the testing stage, with Early Science data already published,
and a Pilot Survey ongoing. The typical ASKAP field of view is $\sim$ 5\fdeg5$\times$5\fdeg5, and the intention is
to do a blind sky survey in the 21-cm HI line of the available sky south of declination +30$^\circ$ (cf. \citealt{koribalski_20}).
A more modest effort is ongoing in the northern sky with the refurbished Westerbork telescope (project 
WSRT-Apertif, \citealt{adams_20}).

The promise of blind HI surveys has already been shown by the ALFALFA survey using the Arecibo telescope,
and recently attention has been given to faint HI-rich ultra-diffuse galaxies, who deviate from the 'standard'
baryonic Tully-Fisher relation (see \citealt{mancera_19}). Such intriguing galaxies, which seem to be baryon
dominated, merit to be studied with CASTLE, so as to obtain much better optical imaging than 
currently available.

\subsection{The case of the Ou4 giant bipolar outflow}
\label{sec:ou4}

Ou4 is a bipolar outflow 
with a total length of 1\hbox{$.\!\!^\circ$}2 on the sky. 
This giant outflow was discovered by the French amateur
astronomer Nicolas Outters while imaging the Sh~2-129 nebula
in June 2011 by means of a 12.5~hour CCD exposure 
with a F/5 106mm-diameter refractor 
(resulting in a pixel scale of 3\hbox{$.\!\!^{\prime\prime}$}5) and a
narrow-band [O\,{\sc iii}]~5007\AA\ filter \citep{acker12}.
\cite{corradi14} obtained shallow broadband $g$ 
and deep narrow-band [O\,{\sc iii}] and H$\alpha$ filter imagery 
of Ou4 at arcsecond resolution
with the 2.5m~INT to study its detailed morphology; 
and long-slit spectroscopy of the tips of the bipolar lobes 
with the 4.2m~WHT to determine the gas ionization mechanism, physical conditions, 
and line-of-sight velocities.
The apparent position and physical properties of Ou4 
are consistent with the hypothesis that Ou4 is located inside the
Sh~2-129 H\,{\sc ii} region, suggesting that it was launched 
by the central, massive triple system. However, the alternate
possibility that Ou4 is a bipolar planetary nebula, or the result of
an eruptive event on a massive AGB or post--AGB star not yet
identified, cannot be ruled out.

\begin{figure}[h]
\includegraphics[width=\columnwidth]{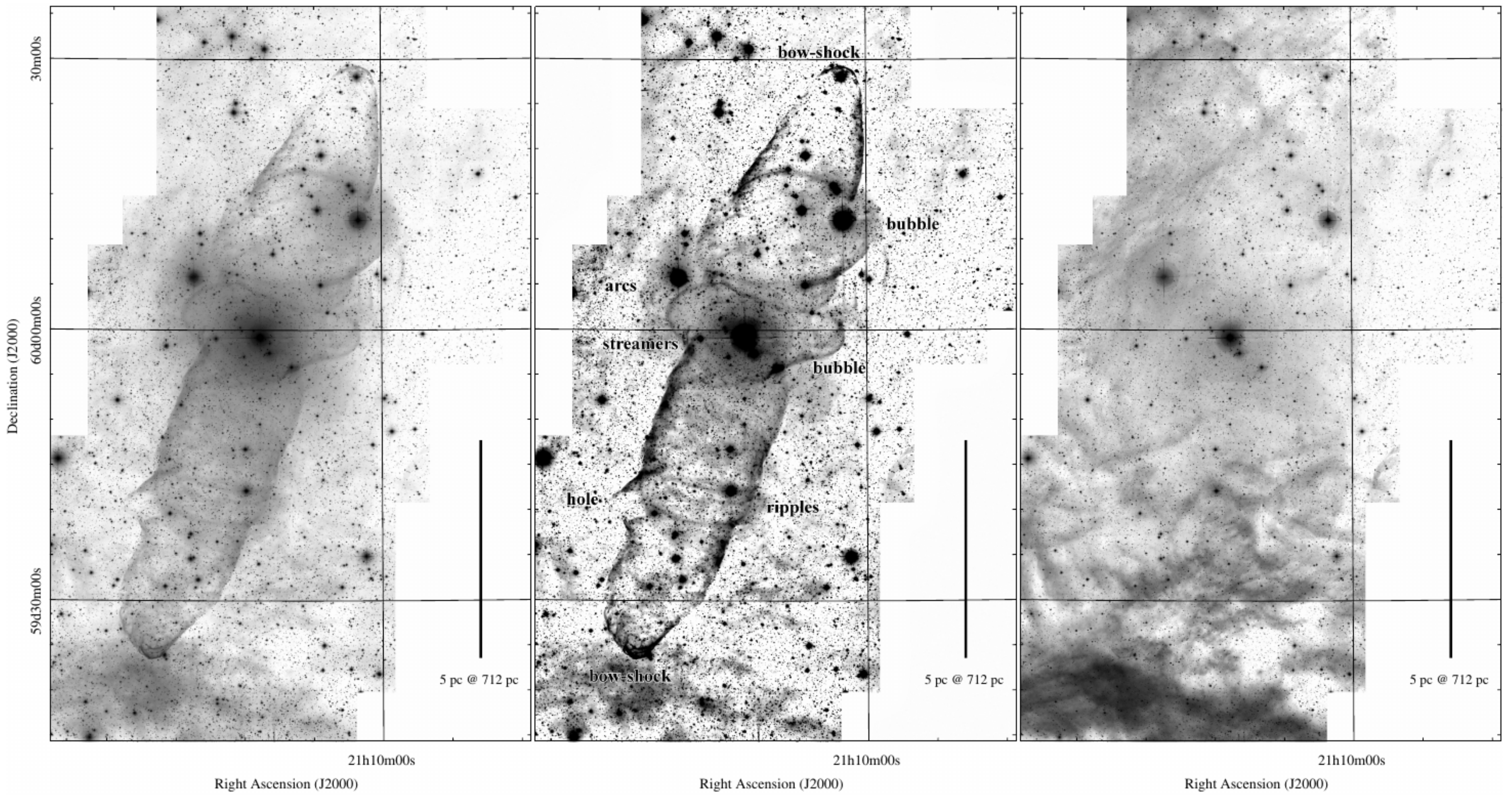}
\caption{[O\,{\sc iii}] (left and middle panels) 
and H$\alpha$ (right panel) INT mosaic images of Ou4 \citep{corradi14}. 
The intensity scale is logarithmic in the left and right panels. 
North is up, east is left.
The mosaic field of view in each panel is 
$\sim$1\hbox{$.\!\!^\circ$}4$\times$$\sim$0\hbox{$.\!\!^\circ$}9, 
which can be surveyed with one CASTLE pointing.
}
\label{figure:INT_mosaics}
\end{figure} 

The complex morphology of Ou4 is mainly visible 
in the [O\,{\sc iii}] INT mosaic 
(left and middle panels of Fig.~\ref{figure:INT_mosaics}): 
multiple bow shocks are detected at the tips of the bipolar lobes, 
in particular the southern one; 
in the central region of the southern lobe a hole is visible, 
as well as ripples directed perpendicularly to its long-axis; 
east of the central distorted bubble, additional features
in the form of streamers and arcs can be identified;
an elliptical bubble breaks the inner regions of the northern lobe. 
The [O\,{\sc iii}] surface brightness ranges from 
$\sim$$4\times10^{-16}$~erg~cm$^{-2}$~s$^{-1}$~arcsec$^{-2}$ 
(tip of southern lobe) down to the detection limit 
of the INT images of several 
$10^{-17}$~erg~cm$^{-2}$~s$^{-1}$~arcsec$^{-2}$, 
which is limited to regions not contaminated by the diffraction halos 
of the bright stars ($V\sim6.6$~mag) at the center of Sh~2-129 
or ghost reflections in the raw images. 
The H$\alpha$ INT mosaic is typically 1.7 times less sensitive 
than the  [O\,{\sc iii}] one.
The H$\alpha$ emission of Ou4 is mainly limited 
to the bow shocks at the 
tips of the bipolar lobes, with a maximum surface brightness of 
$\sim$$3\times10^{-16}$~erg~cm$^{-2}$~s$^{-1}$~arcsec$^{-2}$ 
at the tip of southern lobe; a fainter H$\alpha$ emission 
that is spatially coincident with the [O\,{\sc iii}] emission 
of the northern lobe is also barely detected 
(right panel of Fig.~\ref{figure:INT_mosaics}).

CASTLE with its very low PSF wings and limited ghost 
reflections will improve the imagery of Ou4 in the regions 
where the morphological study is currently limited 
by bright stars; in particular, the regions 
of the central distorted bubble, the arcs, 
and the northern lobe bubble, 
which are critical to constrain the ejection mechanism.
The broadband $g$ filter will select the prominent [O\,{\sc iii}] 
emission from Ou4 and additional emission from 
the H\,{\sc ii} region, which is likely 
faint in the central and northern region of Sh~2-129 
based on the H$\alpha$ INT mosaic
(right panel of Fig.~\ref{figure:INT_mosaics}). 
Therefore, a transformative $g$ image of Ou4 with a 
sensitivity limit (S/N=5 inside 6$^{\prime\prime}$-radius aperture) 
of $10^{-17}$~erg~cm$^{-2}$~s$^{-1}$~arcsec$^{-2}$ 
(corresponding to $V$=28.8~mag~arcsec$^{-2}$) 
could be obtained by CASTLE with an exposure of $\sim$6~h 
(Fig.~\ref{lim_mag}B).

\section{The transient universe}

Astronomy is truly undergoing a revolution in terms of our ability to monitor the time-variability of the Universe in a continuous way using new facilities coupled with advanced machine-learning algorithms.
The opening up of the temporal domain is transforming our knowledge of how the Universe evolves, particularly for objects, which are undergoing explosive change, such as a supernova (SN) or a Gamma-ray Burst (GRB). 
These explosive events can release enormous amounts of power both in electromagnetic radiation and in non-electromagnetic forms such as neutrinos and gravitational-waves, and test our understanding of the laws of physics under the most extreme conditions. 

Observing facilities, which are currently online, enable the sky to be monitored fairly continuously in real-time over large areas and across the electromagnetic spectrum, capturing the temporal behavior of the Universe in a way previously unattainable. 
Examples facilities include the LOFAR radio telescope, the Pan-STARRs telescope and the Zwicky transient facility (ZTF) in the optical and the Swift and Fermi high-energy satellites \citep{harleem_2013,Chambers2016,Smith_2014,Gehrels_2004,meegan_2009}.
Non-electromagnetic facilities are also now observing the transient sky, particularly the Advanced LIGO-VIRGO gravitational-wave observatory \citep{aasi_2015,abbott_2016b,acernese_2015}, which detected in 2015 the first merger from a binary black-hole \citep{Abbott2016} and in 2017 the first coalescence of two neutron stars \citep{Abbott_2017a}, and the Antares, IceCube and Pierre Auger Observatory neutrino experiments \citep{ageron_2011,aartsten_2017,aab_2015,albert_2017}. 
The data from all these facilities have opened up the new {\it Multi-messenger} era, but they are just a foretaste of what is going to come. 
Many of the previously developed theories have come under intense strain by new observational results, such as the highly variable emission seen at late times in GRBs, the discovery of extremely luminous supernovae \citep{Gal-Yam2012} and the still unexplained fast radio bursts (FRB) \citep{Petroff2019}
Theoretical models predict a variety of exotic explosions and stellar mergers, together with their multiple signatures across the electromagnetic spectrum. 
Theory also predicts that some will be accompanied by gravitational wave, neutrino and high-energy particle emission.
In the next decade, the number of transients found will increase by several orders of magnitude as even more powerful facilities come on-line, in particular the Large Synoptic Survey Telescope (LSST), the Square Kilometer Array (SKA), the Space-based multi-band astronomical Variable Objects Monitor (SVOM) and the currently operating eROSITA X-ray \citep{Merloni2012} telescope survey in the electromagnetic domain \citep{robertson_2017,fender_2015,bertrand_2019}. 
This evolution will be accompanied by the rapid progress of all-sky detectors for non-electromagnetic messengers, neutrinos \citep[KM3NeT and IceCube,][]{adrian_martinez_2016,aartsten_2017} and gravitational waves (Advanced-VIRGO,  US and India LIGO, KAGRA, \citep{acernese_2015,LIGO2015,Akutsu2019}), also perfectly suitable for studying the transient sky. 
The next astroparticles detectors and transient sky surveys in the electromagnetic domain will revolutionize our view of the sky by opening new spectral and temporal windows on the Universe. This revolution will lead to many unexpected discoveries. 
However, much of the science extracted from these new windows will require the identification of the electromagnetic counterparts of these multi-messenger transients, as well as their follow-up which would provide important information as the redshift of the corresponding host galaxy and possibly enable their use as standard sirens for cosmology \citep{Abbott17b}.

This quest is not as easy as that. Information delivered is very heterogeneous, the error box varying between several tens of degrees for the astroparticle facilities (as Advanced-VIRGO and LIGO, KM3Net, etc.) to arc-seconds (as for LSST, SKA, etc.).
Difficulty is increased by the expected trigger rate which is very high, in particular if one accepts a reduced trigger threshold to catch the faintest, but also interesting, astrophysical sources (several per night, by integrating over the facilities). 
Recent observations conducted in the framework of the multi-wavelength follow-up of the high-energy neutrinos delivered by ANTARES clearly demonstrated the difficulty of this task. 
The main conclusion of this observation is that the delay between observation and identification must be reduced at maximum (at the minute scale) to trigger further subsequent observations, promoting the development of a fully automatic data processing system.

To meet these challenges, it is necessary to have a telescope with a very large field of view. 
In this aim, CASTLE will cover an important role by specifically addressing the following questions.

\subsection{Gravitational waves}

On September 14, 2015 at 09:50:45 UTC the two detectors of the Laser Interferometer Gravitational-Wave Observatory (LIGO) simultaneously observed a transient gravitational-wave signal \citep{Abbott2016}. 
The signal matches the waveform predicted by general relativity for the merger of a black hole binary \citep{einstein_1916}. 
This first detection opens new prospects and the beginning of an exciting new era of astronomy. 

While an electromagnetic emission is not expected from binary stellar-mass black hole mergers, it is not completely excluded yet. However, one of the most promising sources for joint electromagnetic-gravitational wave observations is coalescing binaries including at least a neutron star. 
Those sources are considered to be good progenitor candidates for GRB of short duration \citep{Eichler1989}.
The first gravitational event of two coalescing binary neutron stars has been detected in 2017 and provides the first direct evidence of a link between these mergers and short gamma-ray bursts \citep{Abbott_2017a}, and was detected by plenty of multi-wavelength ground- and space-based telescopes \citep{Abbott17c}. 

\begin{figure}[htb]
\centering
\includegraphics[scale=0.45]{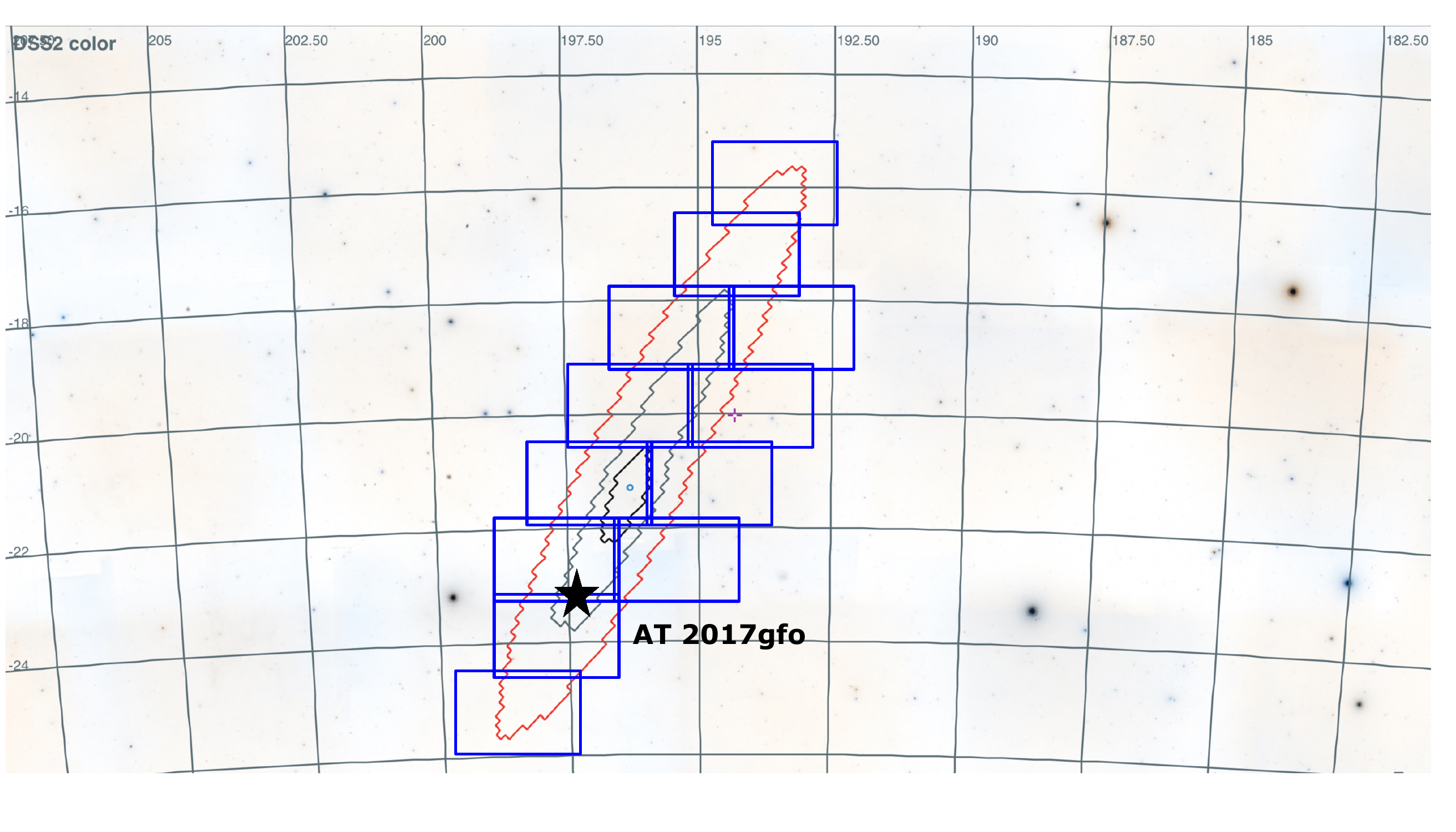}
\caption{The final localization of the BNS gravitational wave event GW170817 as determined by the LIGO-Virgo collaboration (courtesy G. Greco). The black, green and red contour mark the 1-, 2- and 3-$\sigma$ error regions. The blue rectangles correspond to the field of view of CASTLE. to cover the total error region area, CASTLE would need only 12 pointings, which corresponds to a total telescope time of 1.25 hours, including overheads.}
%\index{M83}
\label{AT2017gfo}
\end{figure}

Soon there will be five active gravitational-wave detectors on Earth. 
This worldwide network is expected to observe few to hundreds gravitational-wave events per year associated to neutron-star binary mergers during the future science runs. 
The error regions for gravity-wave transients are typically 50-100 square degree and are usually elongated rather than compact \citep{Abbott2018}. 
CASTLE will be adapted to the monitoring of these alerts and to identify the associated optical counterparts. Its field of view makes it possible to cover the error boxes of the gravitational-wave detectors in several exposures (see example in Figure~\ref{AT2017gfo}), 15 to 30 exposures, while having sufficient sensitivity to probe the volume of Universe covered by them (up-to about 230-250 Mpc, z = 0.05).

\subsection{Neutrinos}
IceCube has demonstrated the existence of neutrinos of astrophysical origin using the outer layer of the IceCube detector as a veto and searching for events starting inside the inner volume \citep{aartsen_2013,aartsen_2013a, aartsen2017} (HESE sample). 
Two event topologies are detected with such detectors: track and cascade event resulting for the muon neutrino charged current interaction and the electron/tau neutrino interaction and muon neutrino neutral current interaction, respectively. Due to the different topologies of the events, the angular resolution is roughly 10\deg-15\deg. for cascades and 0\fdeg5 for muons.
Since 2008, a follow-up of multiplet events, two times per month, is working with optical and X-ray telescope.
The HESE events are now also sent to follow-up facilities through the GCN network \citep{aartsen2017}.
The typical rate of the HESE events is around 5 track events and 10 cascade events per year.
The IceCube Collaboration is planning an expansion of the current detector, IceCube-Gen2 \citep{aartsen_2019}, including the aim of instrumenting a 10 km3 volume of clear glacial ice at the South Pole.

The KM3NeT collaboration is now building the second-generation neutrino telescope in the Mediterranean Sea \citep{adrian_martinez_2016}. 
Above 10 TeV, muon tracks have a typical angular resolution lower than 0\fdeg2 while showers, the most promising events, have a 2\deg localization error. The existing multi-wavelength follow-up program of ANTARES will be extended to KM3NeT.
Ultra-high energy earth-skimming neutrinos can be detected by the large cosmic-ray arrays such as the Pierre Auger Observatory in Argentina \citep{albert_2017} and the large radio neutrino telescope GRAND in China \citep{alvarez_Muniz_2019}. 

Once again, CASTLE will allow the follow-up of these sources without difficulty, knowing that the constraints in terms of field of view and sensitivity are less demanding than in the case of the gravitational-wave detectors. 
Finding the source of these high-energy neutrinos would have a huge impact on the neutrinos-cosmic rays community since they are both related with the acceleration processes in the astrophysical jets.

\subsection{Localization of Fermi Gamma-Ray Bursts}
Gamma-Ray Bursts (GRBs) are the brightest explosions in the universe \citep{Gehrels_2009}. They can be divided into two classes, long and short ones, according to the time during which they emit gamma rays and these classes are believed to correspond to different progenitors and explosion mechanisms \citep{Hjorth_2012,berger_2014,Abbott_2017a}. GRBs are detected by a number of satellites, but NASA's Swift \citep{Gehrels_2004} and Fermi satellites provide the largest number of detections nowadays. To fully exploit these detections, we need imaging and spectroscopic observations of the optical/infrared counterpart to measure the light curve, the redshift, and identify the host galaxies.
The first step in this is a precise localization of the counterpart to within a few arcsec.

The Swift satellite detects about 100 GRBs each year with its wide-field Burst Alert Telescope \citep[BAT,][]{Barthelmy_2005}.
By itself, BAT localizes the GRBs with a precision of about 3\arcmin. The scenario for Fermi/GBM GRBs is very different \citep{meegan_2009, Atwood_2009}. 
Its wide-field gamma-ray burst monitor (GBM) instrument, which covers a wider region of the sky with respect to Swift-BAT, detects about 250 GRBs each year and gives localizations typically of order 100 deg2.

\footnote{\url{https://fermi.gsfc.nasa.gov/ssc/data/analysis/documentation/Cicerone/Cicerone-Introduction/GBM-overview.html}}
Compared to Swift/BAT GRBs, these GRBs are interesting for three reasons.
First, the detection rate is 2.5 times higher. 
Second, the spectral response of Fermi/GBM extends to higher energies than that of Swift/BAT and so it can provide a larger view onto the high-energy prompt spectrum. 
Finally, Fermi/GBM detects about 45 short GRBs per year whereas Swift/BAT detects only about 10 per year, both because if detects more GRBS and since short GRBs tend to be harder and so are missed by the softer response of Swift/BAT\citep{Band2003}. 
The GBM detection rate of short GRBs has recently increased thanks to the use of an offline method that analyses continuous time-tagged event (CTTE) data in order to find sub-threshold short bursts\citep{Kocevski_2018}. Unfortunately, this method can not be executed in real time, due to the limited computational resources available on the spacecraft. 
This type of analysis is more sensitive to burst that are intrinsically weak, or very distant, to trigger directly the GBM detector \citep{Kocevski_2018}.
As an example, the short burst that accompanied the gravitational wave event GW170817A was not an extraordinary event \citep{goldstein_2017} and it was detected at 4.82$\sigma$, although its very low distance (40 Mpc). 
At slightly larger distances, this burst would not have triggered the GBM on-board and it could have been found only after a careful background analysis. The full scientific exploitation of a Fermi/GBM GRB requires localization to within a few arcsec. 

The specificities of CASTLE are perfectly tuned to follow every Fermi/GBM GRB. Our modeling suggests that it will be able to localize about 90 GBM GRBs per year. This will triple the number of precise localizations, from about 20\% to about 60\%, and will significantly advance our ability to effectively study these GRBs. 
Additionally, CASTLE will be on sky roughly at the same time of SVOM which will have a detection rate of about 70 GRB/year \citep{wei_2016} with much smaller error boxes for their localization, of the order of 12\arcmin  to sub-arcsec scale.
CASTLE will be able to detect about 20\% of the alerts in real time and 50\% with 12h delay (day-night effect and rise of the object). 

\section{Upgrades envisioned}

CASTLE will be subject to several upgrades in the future, depending on the community requests and needs. 
\subsection{Stress mirror polishing for primary mirror}
\begin{figure}[htb]
 \begin{center}
  \includegraphics[width=0.49\textwidth]{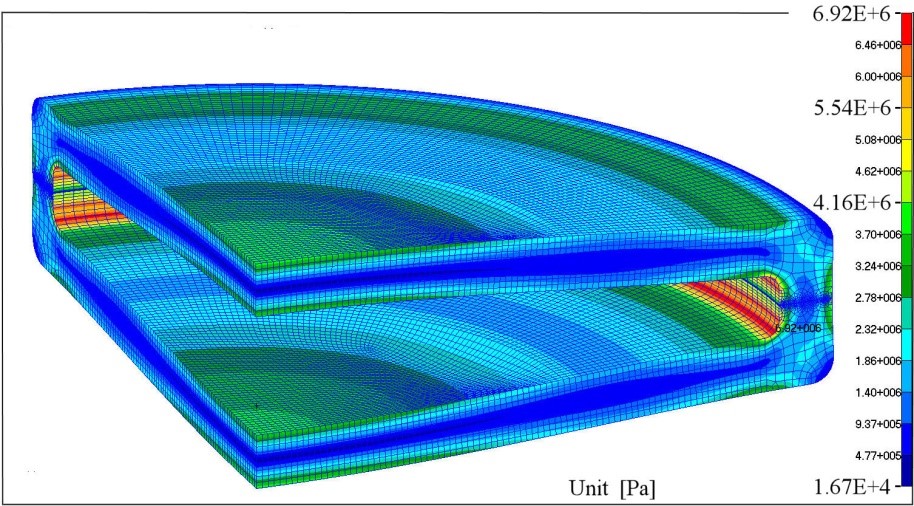}
  \includegraphics[width=0.49\textwidth]{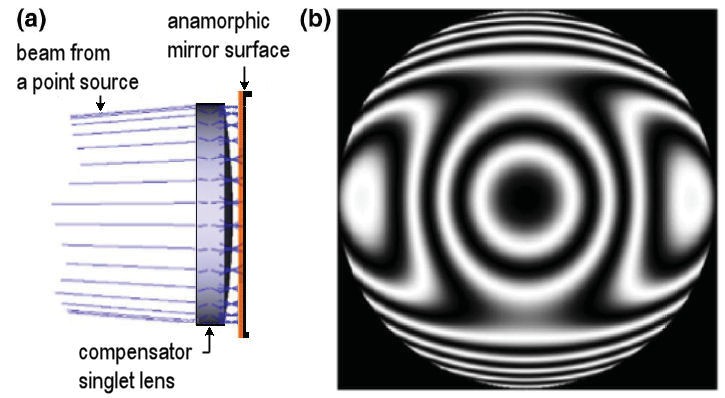}
  \caption{\textit{Left:} stress distribution of the closed vase form during stress polishing. The maximum tensile stress arises along the internal round corners of the elliptical rings with $\sigma$max = 6.92 MPa. For Schott-Zerodur this value is much smaller than the ultimate strength $\sigma$ = 51 MPa for a 1-month loading time duration. \textit{Right:} set-up for testing the freeform primary mirror of CASTLE with null-test singlet-lens aberration compensator. (a) Scheme mounting of the plano-convex lens that compensate for spherical aberration of the primary mirror surface. (b) Simulated He-Ne interferogram of remaining non-axisymmetric fringes to be calibrated \citep{Lemaitre_2018, muslimov}.}
  \label{spm}
 \end{center}
\end{figure}

One of the planned upgrades is the replacement of the monolithic primary mirror with a primary manufactured by using the stress mirror polishing technique (SMP).
Active optics methods \citep{lemaitre} in astronomy provide high imaging quality. 
CASTLE will benefit of highly deformable active optics methods that can generate non-axisymmetric aspheric surfaces -- or freeform surfaces \citep{Forbes_12,hugot_2014} -- by use of a minimum number of actuators. 
The aspheric mirror is obtained from a single uniform load that acts over the whole surface of a closed-form substrate (Figure~\ref{spm}) whilst under axial reaction to its elliptical perimeter ring during spherical polishing \citep{lemaitre_2014,Hugot_08}.

The freeform primary mirror can be generated using the stress mirror polishing technique that provides super-polished freeform surfaces after elastic relaxation. Preliminary analysis required the use of the optics theory of 3rd-order aberrations and elasticity theory of thin elliptical plates. 
Final cross-optimizations have been carried out with Zemax raytracing code and Nastran FEA elasticity code (Figure~\ref{spm}).
This technique allowed determining the complete geometry of a glass ceramic Zerodur deformable substrate, and is expected to be applied to obtain the best image quality, thus providing CASTLE with an extremely high quality primary mirror.
The manufacturing of the primary mirror with the SMP technique will also decrease the large angle scattering of the incoming light, thus limiting its contribution to the PSF wings \citep{Lemaitre_2018}.

\subsection{Filter refurbishing}

\begin{figure}[htb]
 \begin{center}
  \includegraphics[width=0.70\textwidth]{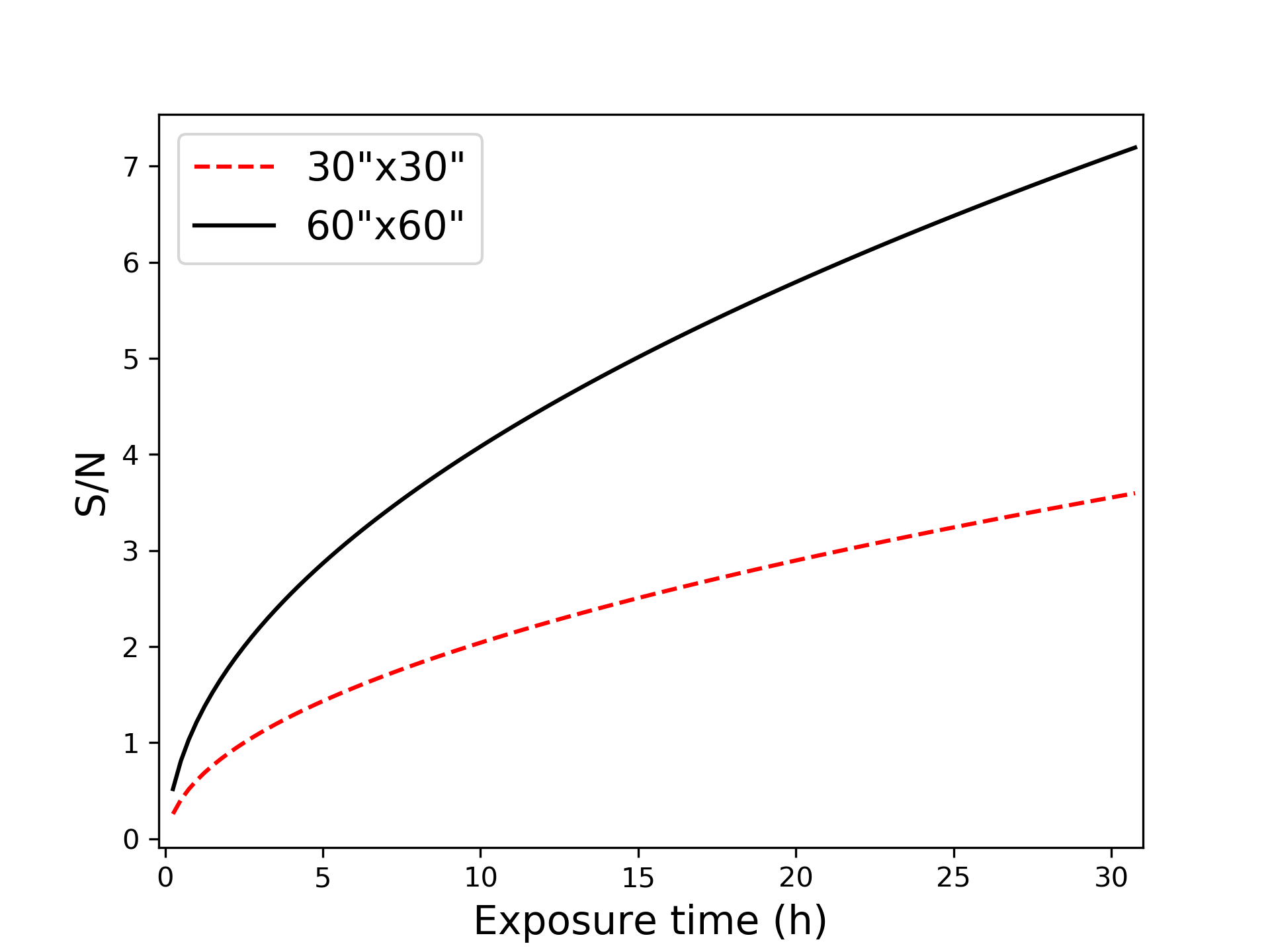}
  \caption{S/N computed for CASTLE when observing with a H$\alpha$ filter and summing the flux over an area of 30\arcsec$\times$30\arcsec (dashed red line) and 60\arcsec$\times$60\arcsec (black line).}
  \label{sn_alpha}
 \end{center}
\end{figure}

Other possible upgrades will concern the filters installed. 
For example we will have the possibility to add an \textit{i}-SDSS filter and a narrow-band filter such as H$\alpha$ or [O III]. 
This could be used to refine some of the science cases and provide deeper images on already visited fields in this band. 
These programs could be carried on specific times during the year, for a fixed amount of time (e.g. two months consecutively), and the filter could be mounted only when needed. 
Since the light collected by the telescope will be greatly reduced when using a narrow band filter, for some of the science cases (eg. the ULSB) it will be necessary to bin over a large number of pixels to increase the S/N. 
In Figure~\ref{sn_alpha} the S/N is shown for CASTLE with a 100\% transmissive H$\alpha$ filter as a function of exposure time for an observed surface brightness of $10^{-18}$~erg~s$^{-1}$cm$^{-2}$arcsec$^{-2}$.

Observations in H$\alpha$ would be extremely valuable for, e.g., the extended tidal features (Section~\ref{sec:tidal_tailes}) and XUV galaxies (Section~\ref{sec:glsb}), since they provide additional information on star forming regions.

Usage of this type of filter would also be important for the imaging of Ou4 (Section~\ref{sec:ou4}).
CASTLE could obtain a deeper H$\alpha$ image than currently available, with the same sensitivity limit as in the g-band observation
(S/N=5 and 11$^{\prime\prime}$ binning) of 
$10^{-17}$~erg~cm$^{-2}$~s$^{-1}$~arcsec$^{-2}$
in $\sim$45~h (Fig.~\ref{sn_alpha}).
Moreover, the visibility of Ou4 at Calar Alto 
is excellent with an airmass lower than 1.5 during 
7~h from mid-August to September~1st. 

\section{Conclusion}
In this paper we showed the main characteristics and performances of the  Calar Alto Schmidt-Lemaitre Telescope (CASTLE) that will be installed in Calar Alto Observatory (CAHA) in 2021/2022. 
Thanks to its unique design CASTLE will be a technological demonstrator for curved detectors and it will be able to provide a wealth of science data and breakthrough discoveries.
In Table~\ref{tab: science_Case} is a list of the different science cases that will be pursued and a possible observation time allocation for each of them.

\begin{table}[ht]
\begin{center}
\caption{Science cases summary and approximate time allocation for observations.}
\begin{tabular}{lr}
\hline\\[-1.5ex]
Science proposed & Available observing time\\
\hline\\[-1.8ex]
Ultra-low surface brightness &   40\%\\ [0.25ex]
Transient search \& detection + solar system & 25\%\\ [0.25ex]
Opened to the community & 20\%\\ [0.25ex]
\hline\\[-1.1ex]
\end{tabular}
\label{tab: science_Case}
\end{center}
\end{table}
CASTLE will observe the ULSB Universe with its large field of view and it will also be sensitive enough to detect transients objects such as GRB, kilonovae and a pletora of explosive events. 
Numerous will be the synergies with ongoing and future surveys (LSST, SKA, SVOM, IceCube, LIGO, VIRGO, etc.), for which CASTLE will provide prompt support to reduce the large error box for source detection and localization and will provide also follow-up when requested.

CASTLE will be additionally well suited for observations of primitive Trans-Neptunian objects (TNO) and Trojans, that are the witnesses of the solar system origin. 
We will be able to observe roughly 24 objects/year with magnitudes in the range 16.5 to 18, by performing observations of 10-30 min per objects with single frame exposure times from 1 to few seconds, due to the fast readout provided by our detector.

Synergies with other stations in Europe, e.g. in Spain (including the Canaries), France or Italy, and northern Africa, may bring important and complementary constraints to determine the shape of the objects, by observing the occultation from other locations.
This will significantly contribute to improve our knowledge on those bodies at a very limited cost.

The possible scientific outcome will not stop there, as there will be many fortuitous observation of transients falling in the large field of view of CASTLE.
These might provide additional time coverage in the transient observation and possible early time data at or before peak luminosity, extremely important for, e.g., reconstruction of GRB and supernova light curves.

\section*{Acknowledgements}
The authors acknowledge the support of the European Research council through the H2020 - ERC-STG-2015 -- 678777 ICARUS program, the MERAC foundation and the European Astronomical Society who are both funding the project. This activity was partially funded by the French Research Agency (ANR) through the LabEx FOCUS ANR-11-LABX-0013. LI acknowledges support from VILLUM FONDEN Investigator grants (projects number 16599 and 25501). LI wishes to thank Giuseppe Greco for his kind availability and support in providing the data presented in Figure 8. JK acknowledges support from NSF through grants AST-1812847 and AST-2006600.
\bibliographystyle{aa} 
\bibliography{main_bib}

\end{document}